\documentclass[aps,preprint, showkeys]{revtex4}%
\usepackage{amsfonts}
\usepackage{amsmath}
\usepackage{amssymb}
\usepackage{graphicx}%
\providecommand{\U}[1]{\protect\rule{.1in}{.1in}}
\begin{document}
\title{ Linked $\mathcal{PT }$-symmetry to Supersymmetry in a class of non-Hermitian Hamiltonians}
\author{Abouzeid. M. Shalaby\thanks{E-mail:amshalab@ mans.edu.eg}}
\affiliation{Physics Department, Faculty of Science, Mansoura University, Egypt}
\begin{abstract}
We introduce and study a class of non-Hermitian Hamiltonians which have
velocity dependent potentials. Since stability can not be advocated directly
from the classical potential, we show that the energy spectra are real and
bounded from below which proves the stability of the spectra of all members in
the class. We find that the introduced class of non-Hermitian Hamiltonians do
have a corresponding superpartner class of non-Hermitian Hamiltonians. We were
able to introduce supercharges which in conjunction with the corresponding
super Hamiltonians constitute a closed super algebra. Among the introduced
Hamiltonians, we show that non-$\mathcal{PT }$-symmetric Hamiltonians can be
transformed into their corresponding superpartner Hamiltonians via a specific
canonical transformation while the $\mathcal{PT }$-symmetric ones failed to be
mapped to their corresponding superpartner Hamiltonians via the same canonical
transformation. Since canonical transformations preserve the spectrum, we
conclude that non-$\mathcal{PT }$-symmetric Hamiltonians out of the introduced
class of Hamiltonians have the same spectrum as the corresponding superpartner
Hamiltonians and thus Susy is broken for such Hamiltonians. This kind of
intertwining of $\mathcal{PT }$-symmetry and Supersymmetry is new as all the
so far discussed cases concentrate on Hamiltonians of broken $\mathcal{PT }%
$-symmetry that have broken Supersymmetry too while we showed that Susy can be
also broken for non-$\mathcal{PT }$-symmetric and non-Hermitian Hamiltonians .

\end{abstract}
\maketitle

\section{Introduction}

Supersymmetry (Susy) plays a very important rule in finding a mathematical
formulation for the interaction of the building blocks of our universe. What
is crucial in the Susy regime is its ability to overcome the famous Hierarchy
puzzle \cite{Hira}. However, the undiscovered super particles predicted by the
Susy theory leaded to the believe that supersymmetry is broken at the current
scale of energy in order to account for the undiscovered super particles.
Since the quantum mechanical case is much simpler to start with for the Susy
breaking, physicists have started to tackle this point in the context of
quantum mechanical models first. In fact, it has been found that the Susy
study in quantum mechanics resembles a very interesting subject by itself. For
instance, one can find the full spectrum of a theory once we know the ground
states of the Hamiltonian operator and its super partner \cite{susy}.
Moreover, for some specific kind of potentials which exhibit shape invariance
one can find the full spectrum as well \cite{shape}.

Susy studies have been applied very recently to the some how new subject of
pseudo-Hermitian Hamiltonians \cite{ptsusy,ptsusy1,ptsusy2,ptsusy3}. However,
in the literature one can find that the studied cases resemble Susy features
of non-Hermitian Hamiltonians of potentials that are position dependent.
Moreover, the connection between $\mathcal{PT }$-symmetry and Supersymmetry in
these studied cases stresses the simultaneous breakdown of both $\mathcal{PT
}$-symmetry and Supersymmetry. In this work, we try to approach the subject
from the side of momentum dependent non-Hermitian potentials and show the
existence of other kind of connection between $\mathcal{PT }$-symmetry and
Supersymmetry in these theories. It is noteworthy to mention that velocity
dependent potentials

Non-Hermitian Hamiltonians with real spectra draw more interest right after
the appearance of the pioneering article of Carl Bender and Stefan Boettcher
\cite{bendr} where they showed that the energy spectra of a class of
non-Hermitian but $\mathcal{PT}$-symmetric Hamiltonians are real and positive.
Their $\mathcal{PT}$-symmetric class has the form;
\begin{equation}
H=p^{2}+x^{2}\left(  ix\right)  ^{n}\text{, \ \ \ }n\geq0\text{.}
\label{qwang}%
\end{equation}
The spectra of such class have been shown, numerically, to be real and
positive \cite{bend1} even in the case of $n=2$. The reality of the spectrum
of such kind of theories is proved to be due to the existence of an unbroken
$\mathcal{PT}$-symmetry for such models. Mostafazadeh showed that the reality
of the spectrum of a Hamiltonian is not limited either to Hermiticity or the
existence of $\mathcal{PT}$-symmetry \cite{spect,spect1}. Instead, he showed
that if a Hamiltonian model $H$ has the property that $\eta_{+}H\eta_{+}%
^{-1}=H^{\dagger}$, then the spectrum of $H$ is real. Here, $\eta_{+}$ is a
Hermitian linear invertible operator and is a positive definite operator as
well wile $H^{\dagger}$ is the Hermitian conjugate of $H$. This formulation of
the problem can be used for Hermitian and $\mathcal{PT}$-symmetric theories as
well as for any pseudo-Hermitian Hamiltonian with respect to a positive
definite metric operator $\eta_{+}$.

In Ref. \cite{ptsusy},\ $\mathcal{PT}$-symmetric theories with possible Susy
structure have been analyzed and shown to be of richer structure than the Susy
Hermitian theories. However, in such theories, the potential is non-Hermitian
and $\mathcal{PT}$-symmetric position dependent operator. In this work,
however, we try to show the intertwining between $\mathcal{PT}$-symmetry and
supersymmetry in non-Hermitian and momentum dependent theories. This case has
not been investigated before and though one can not factorize the Hamiltonian
operator in a normal form that introduce a superpotential, we will show that
the factorization process bears the structure of Supersymmetrey.

Supersymmetry formulation of a theory is closely related to the factorization
of the Hamiltonian operator into the multiplication of two separate operators.
In Hermitian \ Hamiltonians, the factorization of the Hamiltonian takes the
form $H_{1}=A^{\dag}A$, with $A^{\dag}$and $A$ are given by \ $(\hbar=1)$;%

\[
A=\frac{1}{\sqrt{2m}}\frac{d}{dx}+W\left(  x\right)  \text{, \ }A^{\dag
}=-\frac{1}{\sqrt{2m}}\frac{d}{dx}+W\left(  x\right)  \text{\ ,}%
\]
provided that
\[
H_{1}=-\frac{1}{2m}\frac{d^{2}}{dx^{2}}+V\left(  x\right)  ,
\]
where $V\left(  x\right)  $ is the potential. $W\left(  x\right)  $ is called
the superpotential and can be obtained from the Riccati equation of the form;%
\[
V\left(  x\right)  =W^{2}\left(  x\right)  -\frac{1}{\sqrt{2m}}\frac{dW\left(
x\right)  }{dx}.
\]
To construct a Susy theory, one consider the associated Hamiltonian
$H_{2}=AA^{\dag}$. $H_{1}$ and $H_{2}$ have almost the same spectra except
that $H_{1}$ (for instance) has a ground state which is not included in the
spectrum of $H_{2}$. Moreover, the operators $H_{1}$, $H_{2}$, $A^{\dag}$and
$A$ satisfy a graded algebra of the form;
\begin{align}
\left[  H,Q\right]   &  =\left[  H,Q^{\dag}\right]  =0,\nonumber\\
\left\{  Q,Q^{\dag}\right\}   &  =H,\text{ \ }\left\{  Q,Q\right\}  =\left\{
Q^{\dag},Q^{\dag}\right\}  =0,
\end{align}
where $H=\left[
\begin{array}
[c]{cc}%
H_{1} & 0\\
0 & H_{2}%
\end{array}
\right]  $, \ $Q=\left[
\begin{array}
[c]{cc}%
0 & 0\\
A & 0
\end{array}
\right]  $and $Q^{\dag}=\left[
\begin{array}
[c]{cc}%
0 & A^{\dag}\\
0 & 0
\end{array}
\right]  $. In this Susy formulation, the Susy breaking is determined by the
form of the super potential.

In the Susy formulation of non-Hermitian theories one follow a somehow
different algorithm \cite{ptsusy}. The reason is that the Hamiltonian is
non-Hermitian and thus the factorization form $A^{\dag}A$ is no longer
working. However, the Susy breaking determination is not clear yet in the
literature for the Susy formulation of non-Hermitian theories. The point is
that the state functions are not orthogonal and may not be square integrable
in the non-Hermitian representation of a theory. Accordingly, one can not
conclude the Susy breaking even if the ground state is not square integrable
(illusively does not exist). \ For instance, the Hamiltonian of the form
$H=p^{2}+ixp$ is non-Hermitian and has a ground state which is a constant and
thus is not square integrable. However, the corresponding Hermitian
Hamiltonian $h=p^{2}+\frac{1}{4}x^{2}-\frac{1}{2}$ has a square integrable
ground state function ($\psi_{0}\propto e^{-\frac{1}{2}x^{2}}$). In this work
we will introduce a criteria that determines the Susy breaking even in the
non-Hermitian representation of a theory.

The Susy study for the non-Hermitian theories with velocity dependent
potentials has not been stressed yet in the literature. One of the major
problems is that one can not define a super potential for such theories.
However, as we will see later in this work, a factorization of a Hamiltonian
of the form;
\begin{align}
H  &  =H_{0}+gH_{I},\nonumber\\
H_{0}  &  =p^{2},\label{class1}\\
H_{I}  &  =ix^{\epsilon}p,\nonumber
\end{align}
does exist \ and a closed graded algebra can be built. Here, the parameter
$\epsilon$ is a real integer, $x$ is the position operator and $p$ resembles
the momentum operator. Moreover, such Hamiltonians are positive semi-definite.
Since in the class of Hamiltonians in Eq.(\ref{class1}) one can not predict
the boundedness of the spectrum from the Hamiltonian form, a quantum
mechanical study of the boundedness of the spectrum is needed.

The organization of the paper is as follows. In section \ref{semi}, we prove
the positive semi-definiteness of the class of non-Hermitian Hamiltonians in
Eq.(\ref{class1}). The Susy formulation of the class of Hamiltonians in
Eq.(\ref{class1}) will be presented in section \ref{fact}. The breaking of
supersymmetry is discussed in Sec \ref{interplay}. Also, conclusions will
follow in section \ref{conc}.

\section{Positive semi-definiteness of the class of Hamiltonians with
$\mathbf{\mathit{ix}}^{\epsilon}\mathbf{\mathit{p}}$ potentials \label{semi}}

\label{constraint}

A pivotal thing in the supersymmetric formulation of a theory is the existence
of the relation $A\psi_{0}=0$, with $\psi_{0}$ represents the the ground state
function. In other words, for a Hamiltonian to possess a Susy structure it has
to be a positive semi-definite operator. However, the positive
semi-definiteness of the Hamiltonian operators in Eq.(\ref{class1}) can not be
concluded directly from the shape of the potentials. In fact, the class of
Hamiltonians $H$ in Eq.(\ref{class1}) have the same spectra of the equivalent
class of Hermitian Hamiltonians $h=$ $\rho H\rho^{-1}$ , where $\rho
=\sqrt{\eta_{+}}$. Besides, one can easily figure out that $\eta_{+}%
=\exp\left(  -Q\right)  $, where $Q=g\frac{x^{\epsilon+1}}{\epsilon+1}.$ Note
that a state function $\phi_{n}$ in the non-Hermitian representation is mapped
to the state function $\psi_{n}$ in the Hermitian representation via the
relation $\phi_{n}=\rho^{-1}$ $\psi_{n}$. Also, we have the relation
$\rho^{\dagger}=\rho^{-1}$ \cite{spect2}.

To show the positive semi-definiteness of the class of Hamiltonians $H$,
consider the state function;
\begin{equation}
|\chi\rangle=|\chi_{1}\rangle+i|\chi_{2}\rangle,
\end{equation}
with the Hermitian conjugate;
\begin{equation}
\langle\chi|=\langle\chi_{1}|-i\langle\chi_{2}|,
\end{equation}
clearly $\langle\chi|\chi\rangle\geq0$ and thus we get the identity
\begin{equation}
\langle\chi|\chi\rangle=\langle\chi_{1}|\chi_{1}\rangle+\langle\chi_{2}%
|\chi_{2}\rangle-i\langle\chi_{2}|\chi_{1}\rangle+i\langle\chi_{1}|\chi
_{2}\rangle\geq0, \label{ident}%
\end{equation}
or
\begin{equation}
\langle\chi_{1}|\chi_{1}\rangle+\langle\chi_{2}|\chi_{2}\rangle-2Im\langle
\chi_{1}|\chi_{2}\rangle\geq0.
\end{equation}
Now, consider the the expectation value of the Hamiltonian $H$ with respect to
any state function $\phi_{n}$;%
\begin{align}
E_{n}  &  =\langle\phi_{n}\left\vert H\right\vert \phi_{n}\rangle=\langle
\phi_{n}\left\vert \rho^{-1}h\rho\right\vert \phi_{n}\rangle=\langle\rho
^{-1}\psi_{n}\left\vert \rho^{-1}h\rho\right\vert \rho^{-1}\psi_{n}%
\rangle\nonumber\\
&  =\langle\psi_{n}\left\vert h\right\vert \psi_{n}\rangle\nonumber\\
=  &  \langle\psi_{n}\left\vert p^{2}+\frac{1}{4}g^{2}x^{2\epsilon}-\frac
{1}{2}gx^{\epsilon-1}\epsilon\right\vert \psi_{n}\rangle,\\
&  =\langle p\psi_{n}|p\psi_{n}\rangle+\langle\frac{gx^{\epsilon}}{2}\psi
_{n}|\frac{gx^{\epsilon}}{2}\psi_{n}\rangle-\langle\psi_{n}\left\vert \frac
{1}{2}gx^{\epsilon-1}\epsilon\right\vert \psi_{n}\rangle,\nonumber
\end{align}
where we have used the relation $h=$ $\rho H\rho^{-1}=p^{2}+\frac{1}{4}%
g^{2}x^{2\epsilon}-\frac{1}{2}gx^{\epsilon-1}\epsilon.$

But
\[
\frac{1}{2}gx^{\epsilon-1}\epsilon=-i\left[  \frac{gx^{\epsilon}}{2},p\right]
,
\]
then
\begin{align}
E_{n}  &  =\langle p\psi_{n}|p\psi_{n}\rangle+\langle\frac{gx^{\epsilon}}%
{2}\psi_{n}|\frac{gx^{\epsilon}}{2}\psi_{n}\rangle+i\langle\psi_{n}\left\vert
\left[  \frac{gx^{\epsilon}}{2},p\right]  \right\vert \psi_{n}\rangle
,\nonumber\\
&  =\langle p\psi_{n}|p\psi_{n}\rangle+\langle\frac{gx^{\epsilon}}{2}\psi
_{n}|\frac{gx^{\epsilon}}{2}\psi_{n}\rangle+i\langle\frac{gx^{\epsilon}}%
{2}\psi_{n}|p\psi_{n}\rangle-i\langle p\psi_{n}|\frac{gx^{\epsilon}}{2}%
\psi_{n}\rangle,
\end{align}
which has the same form in Eq.(\ref{ident}) with $\chi_{1}=\frac{gx^{\epsilon
}}{2}\psi_{n}$ and $\chi_{2}=p\psi_{n}$. Accordingly, the spectra of the whole
class are all positive and can have the eigen value $E_{0}=0$, which, if it
exists, resembles the ground state of a Hamiltonian $H_{\epsilon}$ out of the class.

Since the square integrability of the ground state functions is pivotal to
determine the Susy breaking, it would be more illustrative to shed light on
their shapes. To do so, let us consider Shr\"{o}dinger equation of the class
in Eq.(\ref{class1});%
\begin{equation}
-\frac{d^{2}\phi_{n}}{dx^{2}}+gx^{\epsilon}\frac{d\phi_{n}}{dx}=E_{n}\phi_{n},
\end{equation}
For the ground state, $E_{0}=0$, we have the solution $\phi_{0}=C$, where $C$
is a constant. Accordingly, the wave function $\psi_{0}$ of the Hermitian
class $h$ is then given by
\begin{equation}
\psi_{0}=\rho\phi_{0}=C\exp\left(  -g\frac{x^{\epsilon+1}}{\left(
\epsilon+1\right)  }\right)  .
\end{equation}
To get the value of the constant $C$ we use;%

\begin{equation}
\langle\psi_{0}|\psi_{0}\rangle=C^{2}\left(  \left(  \frac{(-1)^{\left(
\epsilon+1\right)  }}{\epsilon+1}\right)  ^{-1/\left(  \epsilon+1\right)
}+\left(  \epsilon+1\right)  ^{\frac{1}{\epsilon+1}}\right)  \Gamma\left(
1+\frac{1}{\epsilon+1}\right)  ,
\end{equation}
while the normalized ground state wave function takes the form%

\begin{equation}
\psi_{0}=\left(  \left(  \left(  \frac{(-1)^{\left(  \epsilon+1\right)  }%
}{\epsilon+1}\right)  ^{-1/\left(  \epsilon+1\right)  }+\left(  \epsilon
+1\right)  ^{\frac{1}{\epsilon+1}}\right)  \Gamma\left(  1+\frac{1}%
{\epsilon+1}\right)  \right)  ^{-\frac{1}{2}}\exp\left(  -g\frac
{x^{\epsilon+1}}{\left(  \epsilon+1\right)  }\right)  ,
\end{equation}
We assert that our results are in complete agreement with the analytic
calculations of Ref. \cite{jing}.

\section{Factorization and Susy Algebra of the class $\mathbf{\mathit{ix}%
}^{\epsilon}\mathbf{\mathit{p}}$ of non-Hermitian Hamiltonians \label{fact}}

Let us rename the class of Hamiltonians studied in the previous sections as;%

\begin{equation}
H_{-}=p^{2}+ix^{\epsilon}p.
\end{equation}
An easy realization is \ that the Hamiltonians $H_{-}$ can be factorized as
$H_{-}=ab$, where
\begin{align*}
a &  =-ip+\frac{gx^{\epsilon}}{2},\\
b &  =ip.
\end{align*}
Note that, the factorization $ab$ is different from the factorization
$A^{\dagger}A$ for \ Hermitian Hamiltonians \cite{susy} or the factorization
$A^{\dagger}B$ for $\mathcal{PT}$-symmetric and non-Hermitian theories. The
point is that the class of non-Hermitian Hamiltonians under investigation in
this work possesses $\mathcal{PT}$-symmetric as well as non-$\mathcal{PT}%
$-symmetric Hamiltonians. To show that the Hamiltonians $H_{-}=ab$ in
conjunction with the Hamiltonians $H_{+}=ba$ bears a Susy structure, we
introduce the supercharges $Q_{a}$ and $Q_{b}$ such that;
\begin{equation}
Q_{a}=\left[
\begin{array}
[c]{cc}%
0 & -ip+gx^{\epsilon}\\
0 & 0
\end{array}
\right]  ,\text{ \ \ \ }Q_{b}=\left[
\begin{array}
[c]{cc}%
0 & 0\\
ip & 0
\end{array}
\right]  .
\end{equation}
To investigate the Susy algebra we find that;%
\begin{align}
Q_{b}^{2} &  =\left[
\begin{array}
[c]{cc}%
0 & 0\\
ip & 0
\end{array}
\right]  \left[
\begin{array}
[c]{cc}%
0 & 0\\
ip & 0
\end{array}
\right]  =0,\nonumber\\
Q_{a}^{2} &  =\left[
\begin{array}
[c]{cc}%
0 & -ip+gx^{\epsilon}\\
0 & 0
\end{array}
\right]  \left[
\begin{array}
[c]{cc}%
0 & -ip+gx^{\epsilon}\\
0 & 0
\end{array}
\right]  =0.
\end{align}
Also, the super Hamiltonian $H=\left[
\begin{array}
[c]{cc}%
H_{-} & 0\\
0 & H_{+}%
\end{array}
\right]  $ can be written as;
\begin{align*}
H &  =Q_{b}Q_{a}+Q_{a}Q_{b}=\left[
\begin{array}
[c]{cc}%
0 & 0\\
ip & 0
\end{array}
\right]  \left[
\begin{array}
[c]{cc}%
0 & -ip+gx^{\epsilon}\\
0 & 0
\end{array}
\right]  +\left[
\begin{array}
[c]{cc}%
0 & -ip+gx^{\epsilon}\\
0 & 0
\end{array}
\right]  \left[
\begin{array}
[c]{cc}%
0 & 0\\
ip & 0
\end{array}
\right]  \\
&  =\left[
\begin{array}
[c]{cc}%
p^{2}+igx^{\epsilon}p & 0\\
0 & p^{2}+igpx^{\epsilon}%
\end{array}
\right]  =\left[
\begin{array}
[c]{cc}%
H_{-} & 0\\
0 & H_{+}%
\end{array}
\right]  .
\end{align*}

Note also that;
\begin{align}
\left[  H,Q_{b}\right]   &  =\left[
\begin{array}
[c]{cc}%
p^{2}+igx^{\epsilon}p & 0\\
0 & p^{2}+igpx^{\epsilon}%
\end{array}
\right]  \left[
\begin{array}
[c]{cc}%
0 & 0\\
ip & 0
\end{array}
\right] \nonumber\\
&  -\left[
\begin{array}
[c]{cc}%
0 & 0\\
ip & 0
\end{array}
\right]  \left[
\begin{array}
[c]{cc}%
p^{2}+igx^{\epsilon}p & 0\\
0 & p^{2}+igpx^{\epsilon}%
\end{array}
\right] \nonumber\\
&  =0,\nonumber\\
\left[  H,Q_{a}\right]   &  =\left[
\begin{array}
[c]{cc}%
p^{2}+igx^{\epsilon}p & 0\\
0 & p^{2}+igpx^{\epsilon}%
\end{array}
\right]  \left[
\begin{array}
[c]{cc}%
0 & -ip+gx^{\epsilon}\\
0 & 0
\end{array}
\right] \\
&  -\left[
\begin{array}
[c]{cc}%
0 & -ip+gx^{\epsilon}\\
0 & 0
\end{array}
\right]  \left[
\begin{array}
[c]{cc}%
p^{2}+igx^{\epsilon}p & 0\\
0 & p^{2}+igpx^{\epsilon}%
\end{array}
\right] \nonumber\\
&  =0.\nonumber
\end{align}
Thus the operators $H,Q_{a}$ and $Q_{b}$ constitute a closed superalgebra
$sl(1/1)$.

One can also introduce a super metric operator $\eta=\left[
\begin{array}
[c]{cc}%
\eta_{-} & 0\\
0 & \eta_{+}%
\end{array}
\right]  $ such that

$H^{\dagger}=\eta H\eta^{-1}$. However, for the specific \ class of
Hamiltonians we use, we find that $\eta_{-}=\eta_{+}=\exp\left(
-g\frac{x^{\epsilon+1}}{\epsilon+1}\right)  $.

\section{Relating Susy breaking to $\mathcal{PT}$-symmetry of the class
$\mathbf{\mathit{ix}}^{\epsilon}\mathbf{\mathit{p}}$ of non-Hermitian
Hamiltonians\label{interplay}}

The factorization scheme followed by us for the class of non-Hermitian
Hamiltonians with velocity dependent potentials ($ix^{\epsilon}p$) does not
predict that the supersymmetry of a specific Hamiltonian is either broken or
unbroken. In fact, out of the whole class of Hamiltonians under consideration,
only those of odd $\epsilon$ values are $\mathcal{PT}$-symmetric. Now, it is
very legitimate to make a connection between broken Susy and the
$\mathcal{PT}$-symmetry of a specific Hamiltonian. Note that, since in the
factorization scheme applied above we did not introduce a superpotential, one
can not predict Susy breaking directly from the factorization process. In
fact, in Hermitian and velocity independent supersymmetric Hamiltonians one
can prdict if Susy is broken or not from the shape of the superpotential. For
instance, if the super potential is an odd function in the position operator
$x$ this means that there exists an square integrable ground state function
and thus the theory possesses an unbroken supersymmetry \cite{susy}.

As we do not have a superpotential introduced in our factorization, we seek
another method to predict Susy breaking. For this we assert that Susy breaking
is also characterized by the full equivalence of the spectra of the two
Hamiltonians $H_{-}$ and its its superpartener Hamiltonian $H_{+}$.
Accordingly, if one is able to map the Hamiltonian $H_{-}$ to $H_{+}$ via a
canonical transformation, then Susy is broken for such Hamiltonians. Now
consider the class of non-Hermitian Hamiltonians of the form
\begin{equation}
H_{-}=p^{2}+ix^{\epsilon}p.
\end{equation}

Let us perform the canonical transformation;
\begin{equation}
p\rightarrow-p-igx^{\epsilon},\text{ \ \ }x\rightarrow-x.
\end{equation}
In this case $H_{-}$ transforms as;
\begin{align}
H_{-}  &  =\left(  -p-igx^{\epsilon}\right)  \left(  -p-igx^{\epsilon}\right)
+ig\left(  -x\right)  ^{\epsilon}\left(  -p-igx^{\epsilon}\right) \nonumber\\
&  =p^{2}+igx^{\epsilon}p+igpx^{\epsilon}-g^{2}x^{2\epsilon}-ig\left(
-x\right)  ^{\epsilon}p+g^{2}\left(  -x\right)  ^{\epsilon}x^{\epsilon}.
\end{align}
If $\epsilon$ is an even integer, we find that $H_{-}\rightarrow H_{+}$ which
means that both $H_{-}$ and $H_{+}$ have the same spectra. In other words,
both have the same ground state energy and thus we have \ a broken symmetry.
In fact for even $\epsilon$, the set of Hamiltonians $H_{-}$ is not
$\mathcal{PT}$-symmetric. On the other hand, for odd $\epsilon$ values, the
set $H_{-}$ of Hamiltonians is $\mathcal{PT}$-symmetric. For these cases, the
transformation $p\rightarrow-p-igx^{\epsilon},$ \ \ $x\rightarrow-x,$ does not
map $H_{-}$ to $H_{+}$, which means that $\mathcal{PT}$-symmetric members out
of the class $H_{-}$ may conserve supersymmetry. In fact, this is the case as
one can show that the set of Hamiltonians $H_{-}$ have an equivalent set of
Hermitian Hamiltonians of the form;%
\begin{equation}
h=p^{2}+\frac{1}{4}g^{2}x^{2\epsilon}-\frac{1}{2}gx^{\epsilon-1}\epsilon,
\end{equation}
this Hamiltonian can be written as $h=A^{\dag}A$ where,%
\begin{align}
A  &  =ip+W\left(  x\right)  ,\ \ A^{\dag}=-ip+W\left(  x\right)  ,\nonumber\\
W\left(  x\right)   &  =\frac{gx^{\epsilon}}{2}.
\end{align}
Here $W\left(  x\right)  $ is the super potential corresponding to the
Hamiltonian $h$. Since the ground state function $\psi_{0}$ is given by;
$\psi_{0}=N\exp\left(  -\int W\left(  x\right)  dx\right)  $ \cite{susy} then
for $\epsilon$ odd ($\mathcal{PT}$-symmetric Hamiltonians) $\psi_{0}$ is
square integrable and \ the theory possesses unbroken Susy. On the other hand,
if $\epsilon$ even (non-$\mathcal{PT}$-symmetric Hamiltonians) the
supersymmetry is broken which assures our criteria introduced above for
testing Susy breaking.

\section{Conclusions \label{conc}}

We introduced a non-Hermitian class of Hamiltonians with velocity dependent
potentials of the form $ix^{\epsilon}p$. Though this class have $\mathcal{PT}%
$-symmetric as well as non-$\mathcal{PT}$-symmetric members, the whole class
have real spectra since one can find a positive definite metric operator of
the form $\eta=\exp\left(  -g\frac{x^{\epsilon+1}}{\epsilon+1}\right)  $.
Accordingly, one can find an equivalent Hermitian class of Hamiltonians
$h=\rho H\rho^{-1}$ with $\rho=\sqrt{\eta}$.

We show that the non-Hermitian Hamiltonians $H_{-}=p^{2}+ix^{\epsilon}p$
factorizes as $H_{-}=ab$ with an associated superpartner Hamiltonian
$H_{+}=ba$. For these Hamiltonians we introduced two associated super charges
$Q_{a}$ and $Q_{b}$\ and showed that the operators $H,Q_{a}$ and $Q_{b}$
constitute a closed superalgebra $sl(1/1)$.

In the literature, it has been found that there exists an interplay between
Susy breaking and $\mathcal{PT}$-symmetry breaking \cite{ptsusy}. However, all
super symmetry studies presented in the literature so far stress
$\mathcal{PT}$-symmetric non-Hermitian Hamiltonians with position dependent
potentials. The interplay found in the literature between Susy breaking and
$\mathcal{PT}$-symmetry breaking asserts that theories of broken
$\mathcal{PT}$-symmetry will have associated broken Susy in case of having a
Susy structure. In this work, however, we consider some thing new as we relate
Susy breaking to non-existence of $\mathcal{PT}$-symmetry in non-Hermitian but
velocity dependent potentials. While the class under investigation possesses
theories with exact $\mathcal{PT}$-symmetry, it also includes
non-$\mathcal{PT}$-symmetric members. For members of exact $\mathcal{PT}%
$-symmetry (odd $\epsilon$ values), supersymmetry is conserved but we find
that non- $\mathcal{PT}$-symmetric Hamiltonians out of the class have broken
supersymmetry. This new finding might shed light on the strong interplay
between $\mathcal{PT}$-symmetry and Supersymmetry.

\end{document}